# Scattered light intensity measurements of plasma treated Polydimethylsiloxane films: A measure to detect surface modification.

AMIT R. MORARKA[1*], ADITEE C. JOSHI[1]

[1]Department of Electronics Science, Savitribai Phule Pune University, Pune- 411007, INDIA

[1*] amitmorarka@gmail.com, amitm@electronics.unipune.ac.in

**Abstract.** Polydimethylsiloxane (PDMS) films possess different chemical and physical properties based on surface modification. The bond structure of pristine PDMS films and plasma treated PDMS films differ in a particular region of silicate bonds. We have studied the surface physical properties of pristine PDMS films and plasma treated PDMS films through optical technique. It is already known that plasma treated PDMS films forms very thin $SiO_2$ layer on its surface. Due to difference in coefficient of thermal expansion of the surface $SiO_2$ and the remaining bulk layer, the $SiO_2$ layer develops cracks. These formations are explored to characterize PDMS surface by observing intensity of scattered light while the films are stretched. Pristine PDMS films do not show such optical scattering. Further the intensity measurements were repeated over period of time to monitor surface properties with time. It was observed that plasma treated PDMS film; the scattered light intensity is linearly dependent on the applied force for stretching. After a substantial time, scattering intensity is reduced to the value which is almost equal to that of pristine PDMS surface. It is proposed that basic light scattering through plasma treated PDMS occurs due to the decrease in the width of the cracks when the PDMS film is stretched. The change in the scattering property of plasma treated PDMS surface over time could be attributed to healing of cracks by the migration of polymer chain molecules from the bulk to the surface. It is also reported that with reference to various bonds present on surface plasma treated PDMS surface regains its original properties as that of pristine PDMS with time. Therefore the optical technique could be employed to study surface characteristics of PDMS surface as an alternative approach to conventional spectroscopic techniques.

**Introduction**

PDMS is one of base materials in many microfluidic devices as well as in other domains like bioengineering, flexible electronics and microelectromechanical systems [1]. The material is of prime focus in these domains owing to its characteristics like chemical inertness, ease of fabrication, non- toxicity and low cost. PDMS based applications demand a certain surface properties basically that promotes binding of other layers or molecules for specific applications. Pristine PDMS surface highly hydrophobic that limits further processing of PDMS. In order to modify PDMS surface various techniques have been employed like plasma treatment, physical or chemical treatments, UV irradiation [2-3]. However plasma modification happens to be widely used technique that eliminates use of volatile solvents or hazardous chemical treatments while providing significant modifications on the PDMS surface [3].

But it has been observed that after surface treatment over the time scale the surface regain its properties back to that of pristine PDMS. The surface recovery so far has been detected through IR spectroscopy and adhesion characteristics through contact angle measurements. These methods of surface recovery study someway set some limitations with respect to need of sophisticated instruments like spectrophotometers and cost of systems. The surface recovery study may depend upon availability of instruments and are not cost effective.

In our work, we are devising a novel methodology to detect one such surface property through scattered light intensity measurements. The plasma treated PDMS film shows surface cracks [3-4]. The size of these cracks as measured through an optical compound microscope shows in the range of 1-2 µm. Under zero strain condition, as the light (visual range) is passed through the PDMS film, it does not show cognizable light scattering. As the film is kept under strain, at the same instant scattered light having diffuse intensity is observed. If the light is polychromatic, dispersion of light is observed. To characterize this behavior, a simple mechano-optical setup was developed in-house. As the strain on the film was varied so does the scattered light intensity changed. The intensity of the scattered light also varied with respect to time. To the best of our knowledge, this technique yields better understanding of physical recovery characteristics of PDMS films. The properties of surface modified PDMS films and changes in surface properties over the time period can be analyzed with a better accuracy with time scale as compared to conventional techniques. This may serve as a parallel tool for mapping the surface modification of PDMS.

**Experimental**

**Preparation of PDMS thin films**

PDMS was prepared by mixing Sylgard 184® silicone elastomer and a curing agent (Dow Corning, MI.) in a typical ratio of 10:1 ratio by weight. The mixture was degassed for 20 minutes to remove any air bubbles formed during mixing process. Further, PDMS was spread uniformly on a glass mould and cured at 60° C for three hours. The glass moulds were soaked with ethanol and PDMS thin films were removed and used subsequently.

**Surface Modification**

PDMS thin films were cleaned with deionized water and ethanol respectively. Further, PDMS thin films were inserted in Plasma Asher (EMITECH K1050x) and treated with plasma at 60W for 20 minutes. Plasma treated films were stored at room temperature and inside freezer to check the recovery characteristics after plasma treatment.

**Measurement of scattered light intensity through PDMS thin film**

The scattering intensity was measured by using in-house developed assembly. The schematic of the setup is shown in figure 1. In a typical procedure PDMS thin film was stretched by using a mechanical arrangement and a laser beam was incident on PDMS film. A screen was placed in front of PDMS film which displays scattered light intensities through PDMS film. The scattered light intensity was measured by identifying a diffuse region of scattered light and its intensity was measured by fixing a solar cell exactly at same location.

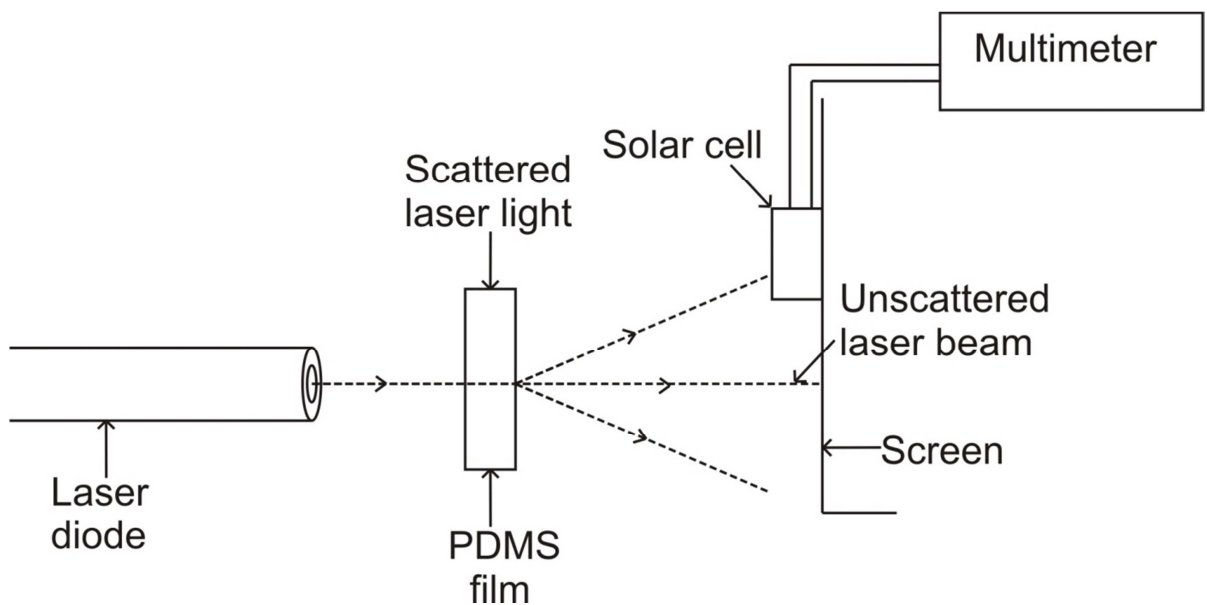

Figure 1: Schematic of setup for scattered light intensity measurement.

The intensity of scattered light was measured by solar cell placed on screen. The PDMS films were stretched by subjecting it to various tensions by added weights. Figure 2 shows arrangement in setup to stretch the film. Stretching was carried out by adding weights in a pan which was attached to a slider-holder mechanism through a metal wire which was routed through a pulley. PDMS film was mounted on to the slider-holder mechanism. It also shows Laser diode and a screen on which solar cell was placed in a region of maximum scattered light intensity.

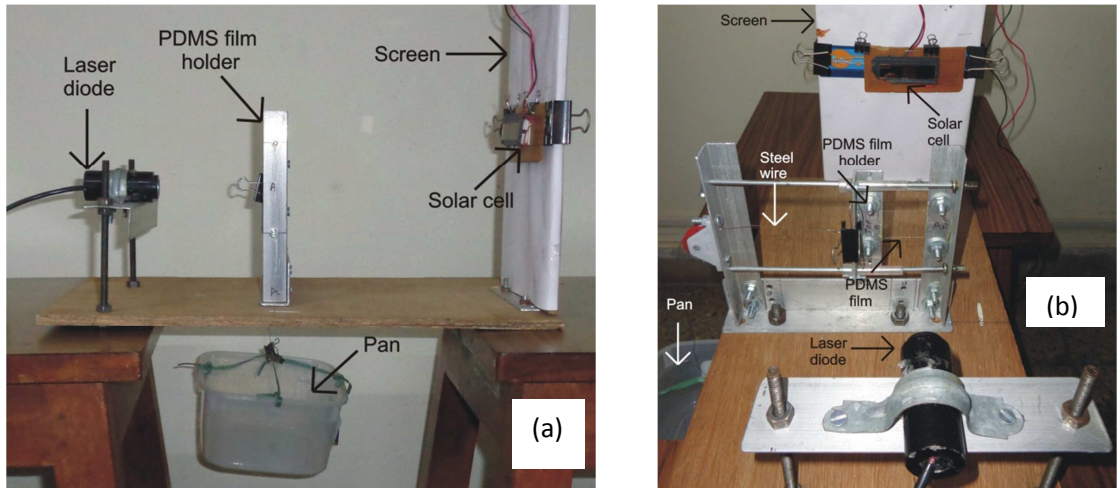

Figure 2: Setup for scattered light intensity measurement. (a) Side View and (b) Top view.

**Results and Discussion**

Figure: 3(a) and 3(b) shows stretched PDMS film with an arbitrarily applied force. In figure 3(a) freshly plasma treated PDMS film shows significant light dispersion (Marked by white rectangle) into its component wavelengths as it was photographed in front of polychromatic light source. The image was obtained by keeping the camera at a tilt angle with respect to the surface of the PDMS film. The same film was used in front of monochromatic light source (laser diode) which shows diffused scattered light projected on to a white screen as shown in figure 4(a) and (b). The image was obtained as mentioned previously. The rectangular hole in the screen of figure 4 is used to channel the unscattered laser beam away from the screen. In all these preliminary experiments it was observed that the light scattering phenomenon through plasma treated PDMS kept under stress, has a spatial symmetry. The figure 4 shows two lobes of scattered light symmetrically situated on either side of the unscattered laser beam.

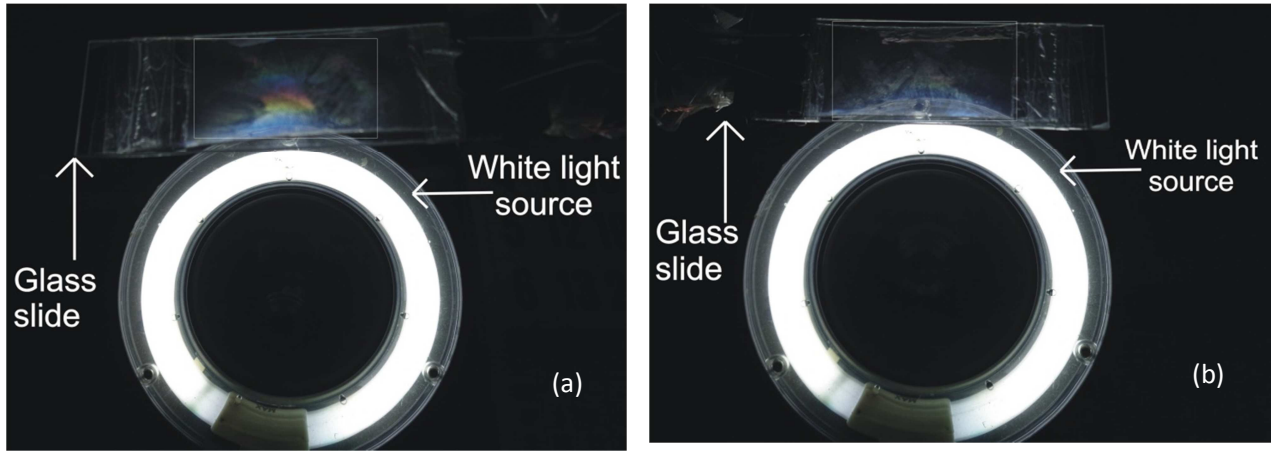

Figure 3: Plasma treated PDMS film subjected to arbitrary force. (a) Freshly plasma treated. (b) After 30 days

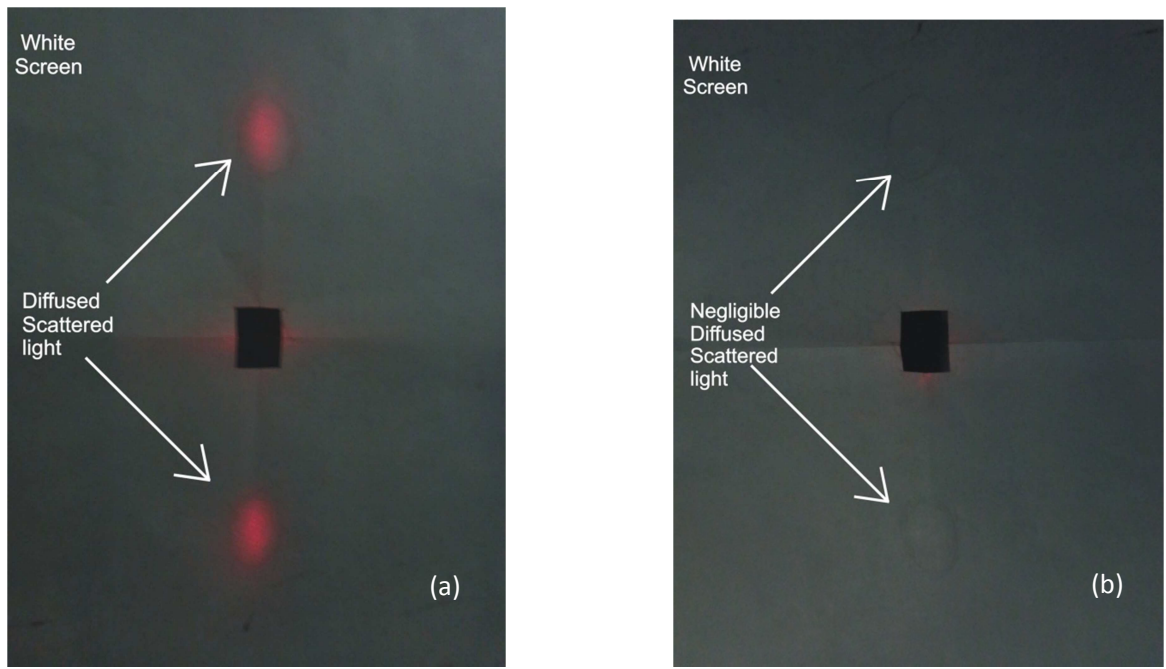

Figure 4: Scattered light intensities of (a) Freshly plasma treated PDMS film and (b) After 30 days.

As reported previously [3-4], plasma treated PDMS films shows surface cracks. The surface of the plasma treated PDMS forms silica like layer [5]. Due to difference in the coefficient of thermal expansion of the bulk PDMS and the silica like layer, results in the formation of cracks. The distribution and orientation of cracks are random in nature over the entire plasma treated surface. The width of these cracks range in 1-2µm [3]. These cracks are responsible for the phenomenon of the scattering of light as mentioned above. We know that for a single slit diffraction to take place, the width of the slit should be comparable to the wavelength of the light. In our case, when the plasma treated PDMS film is exposed to a light source, it does not show any cognizable light scattering. As the same film is stretched (force applied perpendicular with respect to the page of the figure 1), the size of the cracks on the PDMS film changes. This can be seen in figure 5 (a) and (b) respectively. As the applied load is increased on the PDMS film, the increased intensity of the scattered light is shown via increased in the output voltage of solar cell. As the PDMS film is stretched, the orientation of the cracks which lie in the direction parallel to the applied force, their length increases and the width reduces. The width of stretched cracks is less than the unstrained width values (1-2 µm). As such if the magnitude of width approaches to the wavelength of visible light, the cracks act like slits producing the phenomenon of diffraction of light. Since the cracks are orientated randomly on the surface, the diffracted light observed on the screen shows a diffuse lobe like pattern. This is like diffraction of light taking place through randomly orientated slits in one plane. This is also evident from the fact that with the increased applied force on the PDMS film the amount of scattered light intensity is also increased. This happens mainly due to the fact that with increasing strain on the film the width of the cracks decreases which brings their value in the vicinity of the wavelength of visible light. The phenomenon has a symmetrical distribution of scattered light which is similar to that of a single slit diffraction pattern.

**Scattering light intensity measurements**

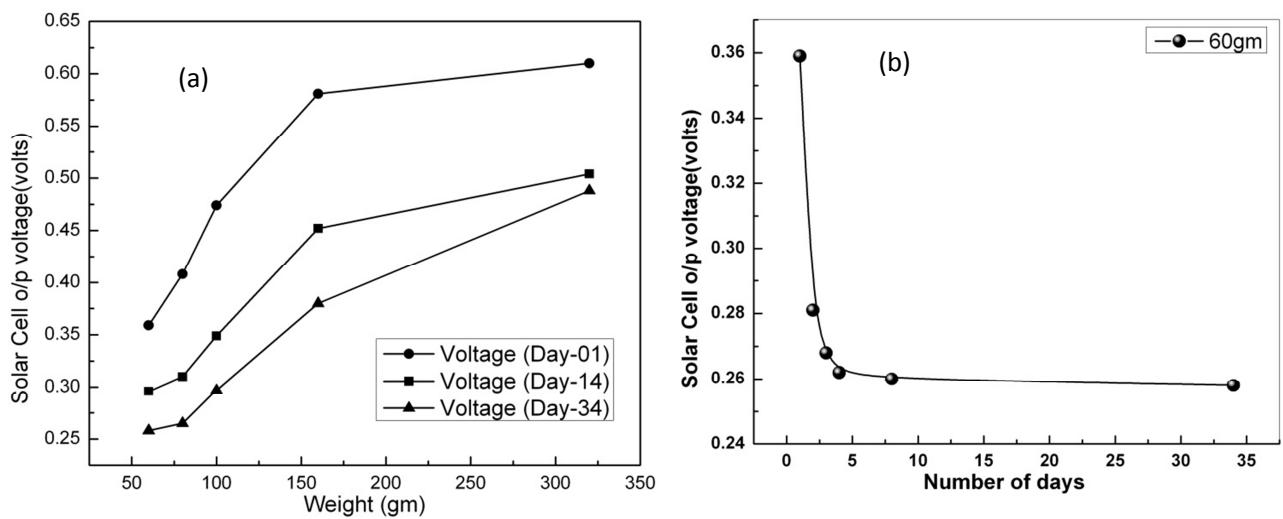

Figure 5: Effect of various loads over the scattered light intensity with respect to time.

As reported by [3] the cracks heal over a period of time. As shown in Figure 5 (a) and (b), the freshly plasma treated PDMS film was fitted in the setup shown in figure 2 and various loads where added in the pan. The graph in figure 5(a) shows the changes in the intensity of light (Changes in the magnitude of voltage produce by the solar cell) over the period of 34 days under the application of various loads. Within the elastic limit of the film, it was stretched to produce maximum intensity of the scattered light. The graph in figure 5(b) shows the changes in the scattered light intensity over the period of 34 days for a load of 60gm. From the two graphs it is evident that the unreacted low molecular weight species from the bulk PDMS film migrates to the surface causing the subsequent loss in the hydrophilicity and the healing of cracks. The plasma treated films which were kept at low temperature showed slow rate of crack healing as compare to the films kept at room temperature.

Thus it was observed that scattered light intensity through plasma treated PDMS film shows gradual decrease with time. The results indicate that the surface recovers back to original characteristics. In order to support our conjecture we have studied FTIR characteristics of PDMS films after plasma treatment and over one month time interval.

**FTIR studies**

PDMS films are hydrophobic in nature. Plasma treatment introduces hydrophilicity on PDMS surface. The PDMS films will regain its properties gradually. These surface characteristics and recovery study of PDMS are reported through FTIR spectroscopy techniques.

In our study, we have taken FTIR spectra of freshly plasma treated PDMS film and over one month time interval. It is reported that plasma treatment introduces functional groups (OH) on surface. Figure 6a shows FTIR spectra for freshly treated PDMS film (curve 'a'), 10 days after plasma treatment (curve 'b') and after one month period (curve 'c'). The spectra show some distinct changes in peak intensity over the time. As seen in figure 6b & 6c, a sharp and intense peak observed at 1258 $cm^{-1}$ and 2963 $cm^{-1}$ in plasma treated PDMS film and the peak intensity observed to be decreased with time. This peak is assigned to asymmetric stretching and deformation of methyl groups Si-$CH_3$ [6]. For freshly plasma treated PDMS film the intensity is high and peak intensity decreases with time, as reported [6-8] plasma treatment introduces OH groups which changes peak intensities for methyl groups. In our results over the time, peak intensity for peaks observed at 1258 $cm^{-1}$ and 2963 $cm^{-1}$, attributed to methyl group gradually decreases indicative of surface recovery of plasma treated PDMS films upon exposure to air with time. This was supported by decrease in intensity for peak at 790 $cm^{-1}$ assigned to Si-C stretching. This is indicative of regaining surface properties with time period. The occurrence of OH group was also evidenced by peaks at 844 $cm^{-1}$ and 3614 $cm^{-1}$ indicative of Si-OH groups [7] on PDMS surface. For this peaks the intensity diminishes with time indicating reduction in OH groups on surface in effect of recovery to pristine PDMS surface.

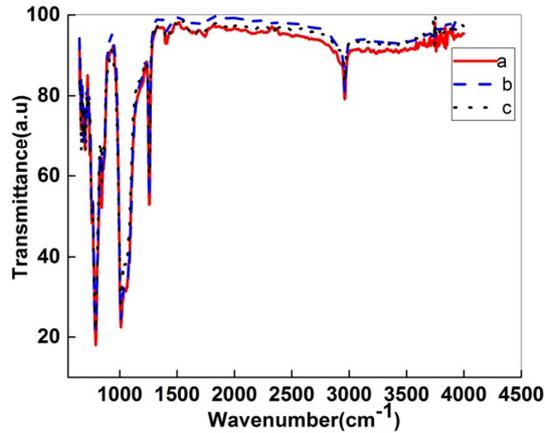

Figure 6a: FTIR spectra of PDMS film. a) Freshly plasma treated PDMS films b) after 10 days c) after one month.

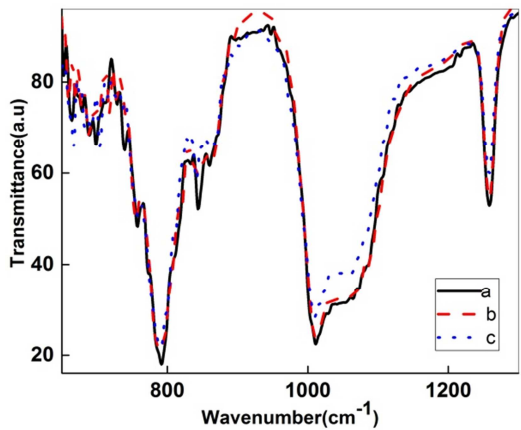

Figure 6b: FTIR spectra of PDMS film in lower range

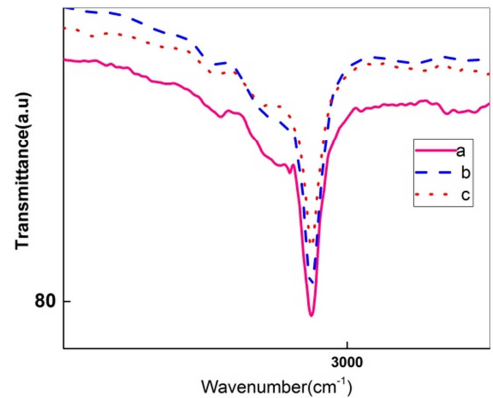

Figure 6c: FTIR spectra of PDMS film in higher range

**Conclusion**

The surface properties of plasma treated PDMS films were studied through optical technique. In this approach surface recovery was studied by measuring intensity of scattered light through PDMS films over one month time period. Plasma treatment creates microcracks on surface of PDMS that acts like slits for scattering light. It was observed that intensity goes on decreasing with time. This is attributed to healing up of microcracks with time reducing amount of light scattered through PDMS films. The surface recovers to hydrophobic properties which were supported through FTIR results. FTIR study of PDMS films over time represents reduction of intensity for hydroxyl bonds indicative of loss of hydrophilicity. Thus scattered light intensity measurement can be used as another tool for mapping of PDMS surface characteristics equally like another spectroscopic measurement techniques.

**Acknowledgement**

The authors wish to thank Prof. S. AnanthaKrishnan, Adjunct Professor and Raja Ramanna Fellow, Department of Electronics Science, Savitribai Phule Pune University (SPPU) for providing his laboratory and the resources for this work. Authors are thankful towards Mr. Somnath Bhopale, Physics department, (SPPU) for his motivational discussions for this work.